\documentclass[aps,prl,nofootinbib,floatfix,twocolumn]{revtex4-2}
\usepackage{amsmath,amssymb,color,graphicx,bm,hyperref,mathrsfs}
\usepackage{physics}
\usepackage{verbatim}

\definecolor{red}{rgb}{0.8,0,0}
\definecolor{RED}{rgb}{0.8,0,0}
\definecolor{violet}{rgb}{0.4,0,0.4}
\definecolor{green}{rgb}{0,0.5,0.0}
\definecolor{GREEN}{rgb}{0,0.5,0.0}
\definecolor{navy}{rgb}{0.0,0.0,0.6}
\definecolor{orange}{rgb}{0.8,0.2,0.0}
\definecolor{blue}{rgb}{0.3,0.0,0.8}
\def\f{\frac}

\begin{document}
\title{\textbf{Laws of thermodynamic equilibrium within first order relativistic hydrodynamics}}
\author{Bhera Ram}
\email{bhera.ram@iitg.ac.in}
\author{Bibhas Ranjan Majhi}
\email{bibhas.majhi@iitg.ac.in}
\affiliation{Department of Physics, Indian Institute of Technology Guwahati, Guwahati 781039, Assam, India.}

\begin{abstract}
Using recently developed consistent and robust first order relativistic hydrodynamics of a dissipative fluid we propose a generalization but weak version of Tolman-Ehrenfest relation and Klein's law on a general background spacetime. These relations are appeared to be a consequence of thermal equilibrium state of the fluid, defined by the absence of heat flux. We interpret them as the defining relations for the local temperature and chemical potential of the fluid. The validity of usual Tolman-Ehrenfest relation and Klein's law deeply depends on the existence of a global timelike Killing vector. However imposition of more stronger equilibrium condition -- local conservation of entropy current -- yields the constancy of the equilibrium thermodynamic parameters along the flow lines.  
\end{abstract}
\maketitle

{\it{Introduction.}}--
Zeroth law of thermodynamics gets modifications in the presence of gravity. Tolman and Ehrenfest \cite{Tolman:1930zza,Tolman:1930ona} showed for static spacetime that a particular quantity $T (x^\alpha)\sqrt{-g_{00}(x^\alpha)}$ remains constant at the thermal equilibrium. This is known as the Tolman-Ehrenfest (TE) relation. Here $T$ is the locally measured temperature of the fluid and $g_{00}$ is the time-time component of the static metric. Later on the same was obtained on a stationary background \cite{PhysRev.76.427.2}. In these analyses, the basic input is the conservation of energy-momentum tensor and the vanishing of the chemical potential $\mu$. However, the non-stationary situations have also been discussed (see e.g. \cite{Santiago:2018lcy}). Consequently using TE relation, Klein \cite{RevModPhys.21.531} showed that $(\mu/T)$ is constant over the spacetime, known as Klein's law. After that, the main focus was to understand the generality and properties of these laws, which led to various approaches to derive them. Cocke \cite{WJ} used the maximum entropy approach which was later followed in \cite{PhysRevD.12.956,Sorkin:1981wd,Gao:2011hh,PhysRevD.85.027503}. A simplification has been done within this approach in \cite{Green:2013ica} by using Noether symmetry formalism. Consequently, an extension of this simplified version to various cases has been followed in \cite{Shi:2021dvd,Shi:2022jya}.  
Use of appropriate thermodynamic ensemble has also been done to obtain them \cite{Roupas:2013fct,Roupas:2014sda,Roupas:2018abp,Roupas:2014hqa}. Rovelli and Smerlek provided an independent concept of ``thermal time'' to get TE relation \cite{Rovelli:2010mv}. Also, see \cite{Gim:2015era} for the generalization of TE at the quantum level.   

Very recently, Lima {\it et al.} \cite{Lima:2019brf,Lima:2021ccv} rejuvenated this paradigm. Using the conservation laws of the energy-momentum tensor (for an ideal fluid), particle number and entropy they argued that in a static spacetime, when $\mu\neq 0$, TE relation and Klein's law are not satisfied separately. Consequently, either of the laws gets modified. Here we aim to investigate this observation {\footnote{A very recent work \cite{Kim:2021kou} in a similar motivation came to our notice. However, our analysis is completely different.}} by adopting a different approach, namely the first-order formalism of relativistic viscous fluid, which has recently been developed in \cite{Kovtun:2019hdm,Hoult:2020eho,PhysRevX.12.021044}. The earlier approaches, by Landau \cite{Landau} and Eckart \cite{Eckart:1940te}, of incorporating dissipation terms at the first order in fluid kinematic and thermodynamics variables within the relativistic regime suffers various issues, like acausality, instability \cite{Hiscock:1983zz}. However later on Israel and Stewart \cite{Israel:1976tn,Israel:1979wp} included higher order terms to build a viable theory of dissipative fluid. In recent years Kovtun \cite{Kovtun:2019hdm,Hoult:2020eho}, adopting derivative expansion technique, has shown that it is possible to construct a consistent fluid description within first order formalism. F. S. Bemfica, M. M. Disconzi and J. Noronha \cite{PhysRevX.12.021044} followed the same, but with different choice of fluid variables, provided a robust and consistent first order relativistic fluid description. We call the later one as BDN formalism. Here we aim to address the TE and Klein relations through this formalism.

It is well known that the usual thermodynamic laws are related to a particular equilibrium condition of the system. For example, the zeroth law is the consequence of thermal equilibrium condition, etc. Therefore it is necessary to investigate those possibilities in order to understand the state of applicability of the two aforesaid laws. We will observe that such not only provides an alternative derivation and generalization of these results, but also illuminates the situation of the system under which  they are viable. 
We find that BDN formalism is very much capable of revealing those important questions. In this formalism one finds various fluid parameters (like viscosity coefficients, heat flux, diffusion flux, thermal conductivity etc.) in terms of thermodynamic quantities (like temperature, chemical potential, etc.) for an arbitrary fluid energy-momentum tensor (in presence of dissipation), provided it satisfies the conservation law as well as consistent with the second law of thermodynamics. Therefore these are capable of defining temperature, chemical potential in terms of fluid parameters. Since the thermodynamic quantities and laws are well defined under the ``equilibrium condition'', we impose respective restrictions on the relations. The exact meaning of equilibrium is being revealed and explained in the main analysis. 

We find that TE relation and Klein's law are coming separately, not like what was shown in \cite{Lima:2019brf,Lima:2021ccv}. Moreover we have been able to provide a generalization of these relations at the thermal equilibrium. Furthermore, we observe that the notion of equilibrium, designated by the absence of any kind of dissipation,  naturally provides the constancy of the fluid thermodynamical variables along the flow lines. 
%Imposing our notion of equilibrium, under any general background we find that the Klein's law is valid all over the spacetime. This is a striking observation as it's validity was derived on a stationary spacetime. 
We also managed to provide a general structure of TE-like relation, which is valid on the hypersurface, orthogonal to fluid velocity. These relations appear to be very fruitful to define local values of temperature and chemical potential. For background with the existence of a global timelike Killing vector this yields the well known TE relation. 

One of the noticeable features of the present analysis is that we have been able to provide generalizations of the two laws. Interestingly it signifies that both the local temperature and chemical potential are individually predicted by the local acceleration of the fluid flow which is generated by the influence of the background spacetime. Moreover these two laws, as expected, seem to be separable under the notion of thermal equilibrium.  Consequently one recovers the usual relations, $T\sqrt{N^2} = $ constant and $\mu\sqrt{N^2} = $ constant over a spacetime, when it has a global timelike Killing vector. Here $N$ relates the $4$-velocity of the fluid and the unit normal to the hypersurface. Since the present analysis is based on BDN formalism, the relation between thermodynamic and fluid parameters are based on the conservation conditions and satisfaction of entropy increase theorem of the fluid. However, unlike in \cite{Lima:2019brf,Lima:2021ccv}, we have been able to discuss the relevant laws in an ``independent'' manner. In addition, contrary to \cite{Lima:2019brf,Lima:2021ccv}, this clearly mention the circumstances under which the laws are satisfied independently.  

Let us now proceed towards the derivation of results. This will justify our aforesaid demands.

{\it{First order formalism.}}--
The first order formalism given by Kovtun \cite{Kovtun:2019hdm,Hoult:2020eho} and then by BDN \cite{PhysRevX.12.021044} is a stable, causal model to study relativistic fluids and takes care of all the short comings of Landau's and Eckart's formalisms \cite{Landau,Eckart:1940te}. The analysis provides a fruitful thermodynamic description of a relativistic fluid in presence of dissipations and viscosity. Here we follow the choice of variables, taken by BDN. Below we give a brief description of the necessary information which are very much needed for our main purpose. However, these are all given in the original work. Therefore almost everything of this section is the repetition of \cite{PhysRevX.12.021044}. In this formalism the fluid description is given as follows.
\begin{comment}
\begin{eqnarray}
p_{\textrm{vis}} &=& - \zeta\Big[\nabla_au^a + \beta_0\dot{p}_{\textrm{vis}} - \alpha_0\nabla_aq^a - \gamma_0Tq^a\nabla_a\left(\f{\alpha_0}{T}\right)
\nonumber
\\
&+& \f{p_{\textrm{vis}}~T}{2}\nabla_a\left(\f{\beta_0u^a}{T}\right)\Big]~; 
\label{tauisr}
\\ 
q^b &=& - \kappa T h^{ab}\Big[\f{\nabla_aT}{T} + \dot{u_a} + \beta_1\dot{q}_a - \alpha_0\nabla_ap_{\textrm{vis}} - \alpha_1\nabla_c\pi^c_a
\nonumber
\\ 
&+& \f{T}{2}q_a\nabla_c\left(\f{\beta_1u^c}{T}\right) -(1 - \gamma_0)p_{\textrm{vis}}T\nabla_a\left(\f{\alpha_0}{T}\right) 
\nonumber
\\
&-& (1-\gamma_1)T\pi^c_a\nabla_c\left(\f{\alpha_1}{T}\right) + \gamma_2 \nabla_{[b}u_{c]}q^c\Big]~;
 \label{heatisr} 
 \\
 \pi_{ab} &=& -2\eta \Big[\beta_2\dot{\pi}_{ab} + \f{T}{2}\pi_{ab}\nabla_i\left(\f{\beta_2u^i}{T}\right) 
 \nonumber 
 \\ 
 &+& \Big<\nabla_au_b - \alpha_1\nabla_aq_b - \gamma_1Tq_a\nabla_b\left(\f{\alpha_1}{T}\right) 
 \nonumber
 \\
 &+& \gamma_3\nabla_{[a}u_{i]}\pi_b^i\Big>\Big]~; 
 \label{tabisr} 
 \\ 
\nu^a &=& - \sigma T^2 h^{ab}\nabla_b\left(\f{\mu}{T}\right)~.
\label{B3}
\end{eqnarray}
\end{comment}
The constitutive relations that give the baryon current and the energy-momentum tensor are 
\begin{eqnarray}
   &&J^a=n u^a~, 
   \label{eq:A1}
\\   
  &&T^{a b}=(\varepsilon+\mathcal{A}) u^a u^b+(p+\Pi) \Delta^{a b}-2 \eta \sigma^{a b}
  \nonumber
  \\
&&+u^a {q}^b+u^b{q}^a~,
   \label{eq:A2}
\end{eqnarray}
where the expressions for $\mathcal{A}$, $\Pi$, $q^a$ and $\sigma^{a b}$ are as follows,
\begin{eqnarray}
&&\mathcal{A}=\tau_{\varepsilon}\left[u^a \nabla_a \varepsilon+(\varepsilon+p) \nabla_a u^a\right]~; 
  \label{eq:A3}
\\
&&\Pi=-\zeta \nabla_a u^a+\tau_p\left[u^a \nabla_a \varepsilon+(\varepsilon+p) \nabla_a u^a\right]~;
    \label{eq:A4}
\\
&&{q}_b =  \sigma T \frac{(\varepsilon+p)}{n} \Delta_b ^a \nabla_a(\mu / T)  +\tau_q\left[(\varepsilon+p) u^a \nabla_a u_b+\Delta_b ^a \nabla_a p\right]~;
     \label{eq:A5}
\\
&&{\sigma}^{a b}=\frac{1}{2}\Delta^{a c}\Delta^{b d}\left( \nabla_c u_d + \nabla_d u_c - \frac{2}{3}\Delta_{c d}\Delta^{e f}\nabla_e u_f \right)~.
     \label{eq:A6}
\end{eqnarray}
\begin{comment}
In the above $\beta_\varepsilon$ and $\beta_n$ are given by
\begin{eqnarray}
&& \beta_{\varepsilon}=\tau_q\left(\frac{\partial p}{\partial \varepsilon}\right)_n+\frac{\sigma T(\varepsilon+p)}{n}\left(\frac{\partial(\mu / T)}{\partial \varepsilon}\right)_n~;
\\
&& \beta_n=\tau_q\left(\frac{\partial p}{\partial n}\right)_{\varepsilon}+\frac{\sigma T(\varepsilon+p)}{n}\left(\frac{\partial(\mu / T)}{\partial n}\right)_{\varepsilon}~.
\end{eqnarray}
Putting the expressions for $\beta_\varepsilon$ and $\beta_n$ into equation~(\ref{eq:A5}), we get a simplified expression for $q_b$ given by
\begin{equation}
{q}_b =  \sigma T \frac{(\varepsilon+p)}{n} \Delta_b ^a \nabla_a(\mu / T)  +\tau_q\left[(\varepsilon+p) u^a \nabla_a u_b+\Delta_b ^a \nabla_a p\right]
\label{eq:A7} 
\end{equation}
\end{comment}
In the above expressions $\varepsilon$, $p$, $n$,$T$ and $\mu$ are equilibrium thermodynamic variables connected via the first law of thermodynamics $\varepsilon + p = T s + \mu n$, where $\varepsilon$ is the equilibrium energy density, $p$ is the equilibrium pressure, $n$ is the equilibrium number density, $T$ is the equilibrium temperature and $\mu$ is the equilibrium chemical potential associated with the conserved baryon charge. Further, $u^a$ is a normalized time-like vector (i.e. $u_a u^a = -1$) called the flow or fluid velocity, and $\Delta_{a b} = g_{a b} + u_a u_b$ is a projector onto the space orthogonal to $u^a$.
Also, $\tau_\varepsilon$, $\tau_p$ and $\tau_q$ quantify the corrections to the out-of-equilibrium contributions to the energy-momentum tensor given by the energy density correction $\mathcal{A}$, the bulk viscous pressure $\Pi$, and the heat flux $q^a$ respectively. Further, $\sigma_{a b}$ is the traceless shear tensor with $\zeta$, $\sigma$, and $\eta$ being the standard transport parameters.

This description is consistent with the required properties \cite{PhysRevX.12.021044}. Particularly the local version of the entropy increase theorem is well satisfied upon using the on-shell conditions $\nabla_a J^a = 0$ and $\nabla_a T^{ab}=0$. The first one yields 
\begin{equation}
u^a\nabla_a n + n\nabla_a u^a = 0~.
    \label{eq:A8}
\end{equation}
The projection of other condition along $u^a$ gives
\begin{eqnarray}
&&u^a\nabla_a \varepsilon + (\varepsilon+p)\nabla_a u^a = - u^a\nabla_a\mathcal{A} - (\mathcal{A}+ \Pi)\nabla_a u^a 
\nonumber
\\
&&\quad - \nabla_a q^a - q^a u^b\nabla_b u_a + 2\eta \sigma_{a b}\sigma^{a b}~; 
\label{eq:A9}
\end{eqnarray}
and its projection on a space perpendicular to $u^a$ is 
\begin{eqnarray}
&&(\varepsilon + p)u^a\nabla_a u^b + \Delta^{a b}\nabla_a p = 
-(\mathcal{A} + \Pi)u^a\nabla_a u^b - \Delta^{a b}\nabla_a \Pi 
\notag 
\\
&&\quad + 2\eta \Delta_c^b(\nabla_a \sigma^{a c}) - u^a\nabla_a q^b  - (\nabla_a u^a) q^b - (\nabla_a u^b ) q^a 
\nonumber\\
&&\quad + u^b q^c u^a\nabla_a u_c~.
\label{eq:A10}
\end{eqnarray}
Use of these on-shell conditions on (\ref{eq:A3}) -- (\ref{eq:A5}) show 
\begin{eqnarray}
&&\mathcal{A}= \order{\partial^2}~;
  \label{eq:A17}
\\
&&\Pi=-\zeta \nabla_a u^a+ \order{\partial^2}~;
    \label{eq:A18}
\\
&&{q}_b = \sigma T \frac{(\varepsilon+p)}{n} \Delta_b ^a \nabla_a(\mu / T) + \order{\partial^2}~.
     \label{eq:A19}
\end{eqnarray}

Next starting with the Euler equation $T s = \varepsilon + p -\mu n$, the covariant version of the Euler relation is given by \cite{PhysRevX.12.021044}
\begin{equation}
T S^a = - T^{a b}u_b + p u^a - \mu J^a~,
    \label{eq:A11}
\end{equation}
where $S^a$ is the entropy current.
On using (\ref{eq:A1}) and (\ref{eq:A2}), we have
\begin{equation}
S^a = s u^a + \frac{q^a}{T} + \mathcal{A}\frac{u^a}{T}~.
   \label{eq:A12}    
\end{equation}
Then the divergence of the above yields
\begin{equation}
\nabla _a S^a = (\nabla_a s) u^a + s\nabla_a u^a+  \nabla_a\left(\frac{q^a}{T}\right) + \nabla_a\left(\mathcal{A}\frac{u^a}{T}\right)~.
   \label{eq:A13}    
\end{equation}
Now using (\ref{eq:A8}), (\ref{eq:A9}), the Euler equation and the law of thermodynamics $Tds = d\varepsilon - \mu dn$, the above equation reduces to
\begin{comment}
 \begin{align}
s\nabla_a u^a = -\frac{(\nabla_a \varepsilon)u^a}{T} - \left(\frac{n \mu}{T}\right)\nabla_a u^a - \frac{u^a \nabla_a\mathcal{A}}{T}\\ \notag
- (\mathcal{A}+\Pi)\frac{\nabla_a u^a}{T} - \frac{\nabla_a q^a}{T} - \frac{q^b u^a\nabla_a u_b}{T} + 2\eta \sigma^{a b}\sigma_{a b}~.
   \label{eq:A13}    
\end{align}   
\end{comment}
\begin{align}
\nabla _a S^a = 2\eta\sigma_{a b}\sigma^{a b} + \mathcal{A}u^a\nabla_a\left(\frac{1}{T}\right) - \Pi\frac{\nabla_a u^a}{T}\\ \notag
- \left(\frac{q^a}{T^2}\right)(T u^b\nabla_b u_a + \nabla_a T)~.
   \label{eq:A14}    
\end{align}
On the other hand use of the Euler equation, the law of thermodynamics and (\ref{eq:A10}) give
\begin{equation}
\Delta^{a b} \nabla_b\left(\frac{\mu}{T}\right) = - \left(\frac{\varepsilon+p}{n T^2}\right)\Delta^{a b}\left[T u^c\nabla_c u_b + \nabla_b(T)\right] + \order{\partial^2}~.
\label{eq:A15} 
\end{equation}
Thus we have 
\begin{align}
\nabla _a S^a &= 2\eta\sigma_{a b}\sigma^{a b} + \mathcal{A}u^a\nabla_a\left(\frac{1}{T}\right) 
- \Pi\frac{\nabla_a u^a}{T} \notag \\
&\quad + \left(\frac{n q^a}{\varepsilon+p}\right)\left[\Delta_a^{ b} \nabla_b\left(\frac{\mu}{T}\right) + \order{\partial^2} \right]~.
\label{eq:A16}    
\end{align}
\begin{comment}
\begin{align}
\nabla _a S^a = 2\eta\sigma_{a b}\sigma^{a b} + \mathcal{A}u^a\nabla_a\left(\frac{1}{T}\right) - \Pi\frac{\nabla_a u^a}{T}\\ \notag
- \left(\frac{n q^a}{\varepsilon+p}\right)\left[\Delta^{a b} \nabla_b\left(\frac{\mu}{T}\right) + \order{\partial^2} \right]
\label{eq:A15}    
\end{align}
    
\end{comment}
Then the out comes (\ref{eq:A17}) -- (\ref{eq:A19}) of the on-shell conditions lead to
\begin{align}
\nabla _a S^a &= 2\eta\sigma_{a b}\sigma^{a b} + \zeta \frac{(\nabla_a u^a)^2}{T} \notag \\
&\quad + \sigma T \left[\Delta^{a b} \nabla_b\left(\frac{\mu}{T}\right)\right]^2
+ \order{\partial^3}~; 
\label{eq:A20}    
\end{align}
which is non-negative when $\eta$,$\zeta$,$\sigma$ $\geq 0$.

{\it Laws in equilibrium state. --}
First we consider that there is no heat flow, i.e. the fluid is at thermal equilibrium. In this case $q^a$=0 and then (\ref{eq:A19}) implies
\begin{equation}
\Delta^{a b} \nabla_b \left(\frac{\mu}{T}\right)= 0~.
    \label{eq:A23}
\end{equation}
In that case Eq. (\ref{eq:A15}) signifies that within the first order formalism we must have
\begin{equation}
\Delta^{a b}\left[T u^c\nabla_c u_b + \nabla_b(T)\right] = 0~.
    \label{eq:A24}
\end{equation}
The equilibrium parameters $T$ and $\mu$, which are connected by the thermodynamic laws, are satisfying these relations.   
Therefore we can interpret them as the defining relations for temperature and chemical potential. Eq. (\ref{eq:A24}) can be expressed in a more convenient form as follows. The left hand side of it reduces to $\Delta^{ab}\nabla_b(\ln T)+\dot{u}^a$ by using the fact $u^a\dot{u}_a=0$ and so $\Delta^{ab}\dot{u}_b = \dot{u}^a$. Here we denote $\dot{u}_a = u^b\nabla_bu_a$, which is the acceleration of the fluid flow. Hence the temperature is identified through the relation
\begin{equation}
\mathscr{D}_a(\ln T)  = -\dot{u}_a~,
\label{B4}
\end{equation}
where $\mathscr{D}_a = \Delta_{ab}\nabla^b$. Since $u^a\mathscr{D}_a=0$, (\ref{B4}) predicts the variation of temperature of the fluid along the spatial directions. Identical relation was obtained in \cite{Santiago:2018lcy}, but considering photon gas. However such was extended to any substance by using few arguments. Here, without using any specific fluid, we have obtained by imposing equilibrium condition on the first order description of fluid. In fact it connects the values of the temperature at different space points of the fluid on a constant time slice. 
Therefore the above one is reminiscent to the TE relation on any general background spacetime at thermal equilibrium. It implies that the local temperature is determined through the acceleration of the fluid which depends on the specific background geometry. Using the above, one indeed can find the exact value of TE expression when the spacetime has global timelike Killing symmetry. We will come back to this later.

Let us now concentrate on (\ref{eq:A23}). This yields 
\begin{equation}
\mathscr{D}_a\Big(\frac{\mu}{T}\Big) = 0~.
\label{B6}
\end{equation}
This again implies that $(\mu/T)$ is constant on the hypersurface $\partial\mathcal{V}$ for any spacetime.  This we interpret as the generalised but weak version of Klein's law.
Further (\ref{B4}) and (\ref{B6}) can be combined to obtain
\begin{equation}
\mathscr{D}_a(\ln\mu) = -\dot{u}_a~.
\label{B7}
\end{equation}
This relation can be treated in similar footing like Eq. (\ref{B4}). It defines the local chemical potential of the fluid.

Now we define a more stronger equilibrium state by the local conservation of the entropy current, i.e. at the equilibrium the fluid constitutive relations satisfy $\nabla_a S^a =0 $ upto second order in derivative expansion. As each term in (\ref{eq:A20}) is positive definite, we must have
\begin{equation}
\sigma^{a b}= 0~; 
%    \label{eq:A21}
%\end{equation}
%\begin{equation}
\nabla_a u^a= 0~;
    \label{eq:A22}
\end{equation}
along with (\ref{eq:A23}) and (\ref{eq:A24}).
Moreover, imposition of these equilibrium conditions along with (\ref{eq:A17}) -- (\ref{eq:A19}) on (\ref{eq:A8}) -- (\ref{eq:A10}) yields the following:
\begin{equation}
(\varepsilon+p)u^a\nabla_a u^b + \Delta^{a b}\nabla_a p = 0~.
    \label{eq:A27}
\end{equation}
and 
\begin{equation}
u^a\nabla_a \varepsilon = 0~;
%    \label{eq:A25}
%\end{equation}
%\begin{equation}
u^a\nabla_a n = 0~;
    \label{eq:A26}
\end{equation}
at the second order.
Note that at the equilibrium all the dissipative terms in the constitutive relations vanish and the equilibrium $T^{ab}$ is defined by quantities like $\varepsilon$, $n$ $s$, $T$, $p$, $\mu$. Therefore we call them as equilibrium parameters of fluid. Moreover the thermodynamics is defined by these variables.   
Then since all the thermodynamic variables depend on $\varepsilon$ and $n$ through the equation of state, one can easily show the following relations:
\begin{equation}
u^a\nabla_a T = 0~;
%    \label{eq:A28}
%\end{equation}
%\begin{equation}
u^a\nabla_a s = 0~;
%    \label{eq:A29}
%\end{equation}
%\begin{equation}
u^a\nabla_a p = 0~;
%    \label{eq:A30}
%\end{equation}
%\begin{equation}
u^a\nabla_a \mu = 0~.
    \label{eq:A31}
\end{equation}
The last set of relations (\ref{eq:A26}) and (\ref{eq:A31}) imply that all the thermodynamic variables ($\varepsilon$, $n$ $s$, $T$, $p$, $\mu$) must be constant along the integral curves of $u^a$. Therefore these fluid thermodynamical variables themselves do not change along the flow lines; i.e. at different time-slices the quantities remain unchanged. Therefore if one assign a timelike coordinate $\lambda$ along $u^a$ such that $u^a=dx^a/d\lambda$, then all these fluid variables are constants of motion. This justifies the demand that at the equilibrium one gets $\nabla_aS^a=0$. The condition $\nabla_au^a=0$ signifies that the flow lines of the fluid do not diverge or contract at the equilibrium. Moreover one can check that in this case the spacetime contains a timelike Killing vector $\xi^a=u^a/T$ such that $\pounds_{\xi}g_{ab} = \nabla_a\xi_b+\nabla_b\xi_a = 0$. This can be checked by using the condition $\sigma_{ab}=0$ and (\ref{eq:A24}). In this case combination of (\ref{B6}) and (\ref{eq:A31}) implies that $\mu/T$ is constant at each event of the spacetime.

We found that (\ref{B4}) and (\ref{B7}) are valid for any type of flow of the fluid, provided there is thermal equilibrium. Therefore the definitions of temperature and chemical potential are very much general and independent of choice of flow. Furthermore these determine the changes of $T$ and $\mu$ with spacetime coordinates in terms of acceleration of fluid. Therefore these temperature and chemical potential gradient can be termed as generalized version of TE relation and Klein's expression for $\mu$. Below we consider a particular type of flow, called as normal flow, to connect our results with the existing ones in literature.  

For generality consider a spacetime whose the metric coefficients are $g_{ab}$. Right now we take $g_{ab}$ as function of all spacetime coordinates. A fundamental observer can be assigned whose four-velocity is $u_a = -N\nabla_a t$, where $t$ is the time coordinate and the hypersurface is one of the $t=$ constant surfaces. Here $u^a$ is normal to the hypersurface and so $\Delta_{ab} = g_{ab}+u_au_b$ is the induced metric. Under these circumstances one can show $\dot{u}_a = \Delta_a^b\nabla_b(\ln\sqrt{N^2}) = \mathscr{D}_a(\ln\sqrt{N^2})$ (see the Eq. ($3.20$) of \cite{Gourgoulhon:2005ng}). Substitution of this in (\ref{B4}) yields $\mathscr{D}_a (\ln T\sqrt{N^2}) = 0$.
Therefore $T\sqrt{N^2} $ is constant on $\partial\mathcal{V}$, but not along $u^a$. This is a weak version of TE relation. Similar was obtained for a few particular spacetimes \cite{Faraoni:2023gqg}. Use of the same argument on (\ref{B7}) yields $\mu\sqrt{N^2}$ is constant on $\partial\mathcal{V}$.
%Now for a choice of time evolution vector field of the form $t^a = (1,N^\alpha)$ one finds $t^a = Nu^a + N^\alpha e_\alpha^a$, where $e^a_\alpha$ is the basis on the hypersurface. In this case $t^2 = -N^2 + N^\alpha N_\alpha$ and so $\mathscr{D}_a (\ln T\sqrt{-t^2+N^\alpha N_\alpha}) = 0$. Identical expression exists for chemical potential as well.
However for a spacetime with a global timelike Killing vector, $T\sqrt{N^2}$ and $\mu\sqrt{N^2}$ do not change with time. In this case  $T\sqrt{N^2}$ and $\mu\sqrt{N^2}$ are constant all over the spacetime. This has been obtained in \cite{Santiago:2018lcy} by considering a stationary spacetime. These will lead to the usual relations when $u^a$ is chosen along the timelike Killing direction corresponding to time translation; i.e. the fluid performs Killing flow along time coordinate \cite{Santiago:2018lcy}.

{\it{Conclusions.}}--
Within the first order formalism, given in \cite{PhysRevX.12.021044}, we found generalised versions of TE relation and Klein's law. Various important features can be noted here. These are as follows. 
\begin{enumerate}
\item The analysis is being done within the first order formalism of dissipative fluid. 
\item At the thermal equilibrium local temperature and chemical potential are determined through the acceleration of the fluid (see Eq. (\ref{B4}) and Eq. (\ref{B7})). 
\item These relations, contrary to earlier attempt \cite{Santiago:2018lcy}, are derived without invoking any specific type of fluid. Therefore generality of them is quite vivid. 
\item $T\sqrt{N^2}$ and $\mu\sqrt{N^2}$ are constant on the hypersurface. The first one can be interpreted as the weak version of TE relation. These are constant at each spacetime event when the background bears a global timelike Killing vector. 
\item At this equilibrium $\mu/T $ is constant on the hypersurface for any spacetime, a weak version of Klein's law. 
\item Contrary to \cite{Lima:2019brf,Lima:2021ccv}, here both the required relations are coming separately. 
\item At the equilibrium where there is no-dissipation, all the thermodynamic parameters are constant along the fluid flow lines. In this case the spacetime metric carries a timelike Killing vector.
\end{enumerate}

%The most striking and surprising observation is -- the quantity $(\mu/T)$ is constant at each spacetime event for any general background metric. So the domain of validity of the Klein's law seems to be much wider than it's earlier prediction. However this analysis is within the first order description of fluid. Therefore it needs to be further investigated to know the robustness of such claim.  

Finally, we want to mention that all these results and conclusions were drawn within the first order formalism by BDN. Hence one must be wonder whether the same also hold for other formalisms to study the thermodynamics of a relativistic fluid. Whatever be the situation, since BDN formulation is causal and free of any ambiguity, we feel that the overall conclusion must be universal. But to obtain a conclusive evidence it is necessary to perform further investigation.

\vskip 3mm
\begin{acknowledgments}
{\it Acknowledgments.}-- The work of BR is supported by the University Grants Commission (UGC), Government of India, under the scheme Junior Research Fellowship (JRF). BRM is supported by Science and Engineering Research Board (SERB), Department of Science $\&$ Technology (DST), Government of India, under the scheme Core Research Grant (File no. CRG/2020/000616). 
%Both the authors like to thank the anonymous referee for pointing out few issues of the first version which led to improvement of the results.
\end{acknowledgments}

\bibliographystyle{apsrev}

\bibliography{bibtexfile}

\end{document}